\documentclass[journal,12pt,onecolumn]{IEEEtran}

\usepackage{epsfig}
\usepackage{algorithm}
\usepackage{algorithmic}

\usepackage{fullpage}
\usepackage{cite}
\usepackage[cmex10]{amsmath}
\usepackage{algorithmic}
\usepackage{mdwmath}
\usepackage{mdwtab}
\usepackage[tight,footnotesize]{subfigure}

\newcommand{\qed}{\nobreak \ifvmode \relax \else
      \ifdim\lastskip<1.5em \hskip-\lastskip
      \hskip1.5em plus0em minus0.5em \fi \nobreak
      \vrule height0.75em width0.5em depth0.25em\fi}

\begin{document}

\title{A new algorithm for extracting a small representative subgraph from a very
  large graph}
\author{\IEEEauthorblockN{Harish Sethu and Xiaoyu Chu}\\
\IEEEauthorblockA{Department of Electrical and Computer Engineering\\
Drexel University\\
Philadelphia, PA 19104-2875\\
Email: \{sethu, xiaoyu.chu\}@drexel.edu}
}

\maketitle
\thispagestyle{empty}
~\vskip 0.5in
\begin{abstract}
Many real-world networks are prohibitively large for data retrieval,
storage and analysis of all of its nodes and links. Understanding the
structure and dynamics of these networks entails creating a smaller
representative sample of the full graph while preserving its
relevant topological properties. In this report, we show that graph sampling 
algorithms currently proposed in the literature are not able to
preserve network properties even with sample sizes containing as many
as 20\% of the nodes from the original graph. We present a new sampling
algorithm, called Tiny Sample Extractor, with a new goal of a sample
size smaller than 5\% of the original graph while preserving 
two key properties of a network, the degree distribution and its
clustering co-efficient. Our approach is based on a new empirical method of
estimating measurement biases in crawling algorithms and compensating
for them accordingly. We present a detailed comparison of best known
graph sampling algorithms, focusing in particular on how the
properties of the sample subgraphs converge to those of the original
graph as they grow. These results show that our sampling algorithm
extracts a smaller subgraph than other algorithms while also achieving
a closer convergence to the degree distribution, measured by the
degree exponent, of the original graph. The subgraph generated by the
Tiny Sample Extractor, however, is not necessarily representative of
the full graph with regard to other properties such as
assortativity. This indicates that the problem of extracting a truly
representative small subgraph from a large graph remains unsolved.
\end{abstract}

\newpage
\section{Introduction}
\label{sec:intro}

Large networks, usually modeled as graphs, appear in a variety of
contexts in computer science as well as in sociology, epidemiology,
business and engineering
\cite{New2003a}. Within
computer science, tools that give us insight into the structure and
dynamics of these networks are central to understanding the growth and
evolution of the Internet \cite{SigFal2003}, the nature of online
social interactions \cite{MisMar2007,KumNov2006}, data sharing patterns on peer-to-peer
networks \cite{StuRej2006}, and online epidemic behaviors (whether of ideas
\cite{KemKle2003,GoeMyn2004,CarNag2007} or computer viruses and worms
\cite{OdoSet2004}). Some of these networks, however, are so
large that technical limitations of storage, computing power, and
bandwidth available to most researchers make it infeasible to crawl
through the entire network (e.g., YouTube with over hundred million
nodes or the network of web pages with billions of nodes). Collection of
temporal data for understanding the evolution of these networks
further increases the challenge because of the need for multiple
snapshots of the networks. Even if the data is acquired, they can be
prohibitively large for purposes of analysis, simulation or
visualization on most computing systems. These challenges call for a
fast algorithm that visits only a small fraction of a large graph to
extract a sample subgraph which retains the most important topological
properties of the original graph. The size of the sample subgraph
needs to be significantly smaller than the original graph and free of
measurement bias \cite{LakBye2003}.

As we show later in this report, currently known sampling algorithms do
not quite approach the properties of the original graph even with
sample sizes as large as 20\%. In this work, we pursue the above
challenge with a target of shrinking a network to less than {\em five
  percent} of its original size while preserving a key property of the
graph, its degree distribution. As has been argued in
\cite{HubKri2008}, finding the optimal subgraph $S$ of a certain size
that best matches a property of the original graph $G$ is an
NP-complete problem for most graph properties. Given that the size of
$G$ is often of the order of tens of millions, finding the optimal
subgraph is obviously not feasible. A further constraint that adds to
the challenge is the fact that all of the graph $G$ is not usually
visible to the crawler or is not even accessible. This is frequently
the case while crawling an online social network, the network of web
pages or peer-to-peer networks.  Often, however, if access is secured
to a node, it is possible to secure access to the neighbors of the
node. The goal is to extract a representative subgraph based on
crawling through the original graph beginning with a node in a known
portion of the  graph. A further goal is that the sampling algorithm
be scalable, i.e., its computational costs should increase linearly
with the desired size of the sample subgraph and not depend upon the
size of the original graph.

In the following, we now formalize our problem statement. Consider a
large graph $G$ of $n$ nodes. Our goal is to extract a subgraph $S$ of
$G$ with the following properties:  
\begin{enumerate}
\itemsep -2pt
\item $S$ has $h$ nodes where $h \leq 0.05n$.
\item $S$ has a degree exponent as close as possible to the original graph $G$.
\end{enumerate}
under the following constraints:
\begin{enumerate}
\itemsep -2pt
\item The number of nodes of $G$ visited by the sampling algorithm is
  $O(h)$.
\item Properties of the original graph $G$ are not inputs to the sampling algorithm.
\end{enumerate}

Section \ref{sec:related} presents known solutions related to the
problem statement described above and builds the rationale for the
algorithm proposed in this report. Section \ref{sec:biased} presents our
{\em Tiny Sample Extractor}, a sampling algorithm that uses a biased
random walk to discover new nodes and returns to the starting point
upon discovery of each new node. The bias is used to compensate for
the skewed distribution of nodes visited in random walks or a
breadth-first search. Section \ref{sec:results} presents a comparison
of our approach to other sampling strategies with respect to a number
of graph properties including degree distribution, assortativity and
clustering properties. The results show that our algorithm is able to
extract a much smaller representative sample than other sampling
algorithms while also achieving a closer convergence to the degree
distribution of the original graph. Our results also show that a
sample generated for preserving a particular property can fail to
adequately preserve another property. Section \ref{sec:conclusion}
concludes the report.

\section{Related Work}
\label{sec:related}

A variety of strategies have been used to ``shrink'' a graph for
purposes of analysis. Shrinking a graph by extracting an actual
subgraph allows one to discover patterns in the subgraph that can be
validated later in the original graph because of the one-to-one
correspondence between the nodes in the two graphs. The generation of
synthetic topologies which have a specific set of properties in them
\cite{AirCar2005,EubKum2004}, though useful in many contexts, is not
considered in this report.

A sample subgraph induced by a randomly selected set of nodes has been
discussed in several works on graph sampling
\cite{BecCas2006,Ben2007,HubKri2008,ShiBon2008}. Selecting nodes randomly ensures
that nodes of a given degree are chosen 
with probability proportional to the number of such nodes in the
network. The selected set of nodes have a degree 
distribution very similar to that of the original graph, but these
degrees are the degrees of the nodes in the original graph $G$ and not
the degrees in the induced subgraph $S$. There are at least
two additional problems with such a sampling strategy: (i) when the desired
sample size is as small as 5\% of the original graph, the induced
subgraph is highly likely to be a disconnected graph even if $G$
is connected and thus, unrepresentative; and (ii) in real networks
that have to be crawled, it is usually very hard or infeasible to
generate a statistically valid set of uncorrelated random nodes from
the full graph $G$ given that the full graph is not known (even though
a few random nodes can always be selected from within the known
portion of the graph).  

A related set of sampling strategies is based on selecting 
random edges instead of random nodes or a combination of node
and edge sampling \cite{Ben2007}. In general, however, edge
sampling does not overcome the problems of node sampling mentioned
above. Sampling strategies based on random deletion
\cite{KriFal2005,KriFal2007} instead of selection also suffer the
same problems and are not suitable as solutions to the problem
statement expressed in Section \ref{sec:intro}. Node or edge sampling 
is useful in contexts where the goal is to infer properties of nodes
but not necessarily the topological properties of the graph. The choice
of nodes guided by simulated annealing can target a specific set of
topological properties \cite{HubKri2008}, but this method also relies
on randomly choosing nodes from the entire network.

As an improvement, one may resort to a random walk on the graph to
ensure that a connected set of nodes is chosen for the sample. As
discussed in \cite{StuRej2006}, however, a random walk visits a node
with probability proportional to its degree, leading to a biased
sample\footnote{The random walk technique is identical to one where we
select nodes at random with a probability proportional to its PageRank
\cite{BriPag1998}.}. This is corrected in the {\em
  Metropolized Random Walk (MRW)} \cite{StuRej2006}, based on the Metropolis-Hastings method
for Markov chains \cite{ChiGre1995}. In MRW, a move from node $x$ to node $y$ is
made with probability $P(x,y)$ given by:
\[
P(x,y) = \frac{1}{\mbox{degree}(x)}
\]
and then, the move is accepted with probability:
\[
\displaystyle{ \mbox{min} \left( 1, \frac{\mbox{degree}(x)}{\mbox{degree}(y)} \right) }.
\]
If the move is not accepted, we return to node $x$ and attempt a move
again. The expected distribution of the degrees (in graph $G$) of the
visited nodes in MRW is identical to the actual distribution of the
degrees in $G$. However, the subgraph induced by the set of visited
nodes is highly unlikely to have a similar distribution as $G$. This
is because a random walk is more likely to yield a ``string'' of connected nodes rather
than a scaled-down network. The MRW algorithm, however,
is used in our approach, not to directly extract a sample subgraph but
to first obtain an estimate of the degree distribution in the original
graph $G$. 

A further improvement in graph sampling is achieved with {\em Snowball
  sampling}, which chooses a random node from the known portion of the
graph $G$ and then proceeds with a breadth-first search until the
desired size of the sample graph is achieved \cite{New2003b}. Snowball
sampling and its derivatives have been used in 
social network analysis \cite{LeeKim2006,MisMar2007}. As reported in
\cite{AhnHan2007}, for small sample sizes, it is inconclusive if the
clustering co-efficient of the sample network converges to that of the
complete network (as will be verified in our work as well). In
addition, Snowball sampling has been shown to over-sample ``hubs'' or
large-degree nodes in a network because of its breadth-first strategy
which hits a hub with a greater likelihood. 

A related strategy is one called {\em Forest Fire}, first introduced
in \cite{LesKle2005}. In this method, as in Snowball sampling,
we choose a random node from the known part of $G$ and use a
breadth-first approach. With a ``forward burning probability'' $p_f$,
the node burns links attached to it. The nodes at the other end of a
burned link are added to the sample subgraph and they now continue
spreading the ``fire'' by burning links attached to them. This
continues until the desired size of the sample subgraph is achieved. 
It has been found in \cite{LesFal2006} that the Forest Fire sampling
strategy works best with $p_f = 0.7$ and this is what we use in all our
simulations in this report. In general, it has been found that methods
based on BFS search are likely to overestimate node-degrees and
underestimate symmetry \cite{LeeKim2006}. As we will show later, the
Forest Fire sampling strategy, being based on a ``scaled-down'' BFS,
is not entirely able to reduce the likelihood of adding high-degree
nodes to the sample subgraph.

A more rigorous but different approach to random subgraph sampling has
only recently been attempted in \cite{LuBre2012} which evaluates a number of
different strategies including Random Vertex Expansion
\cite{KasItz2004}. However, while the subgraphs sampled in the methods
proposed in \cite{LuBre2012} achieve a sampling of subgraphs uniformly
at random, they do not actually 
extract a single subgraph that is most representative of the full graph
with respect to any given property. As a result, random sampling of
subgraphs do not readily help us discern properties of the full graph,
especially since the sampling of subgraphs uniformly at random leads
to an over-representation of properties from dense portions of the
graph. 

\section{The Tiny Sample Extractor}
\label{sec:biased}

Our algorithm relies on first finding an estimate of the degree
distribution of nodes in the original graph $G$. We use the
Metropolized Random Walk for this purpose and use the degree exponent,
$\cal{D}$, as defined in \cite{SigFal2003}, to capture the degree
distribution. The complementary cumulative distribution function
(CCDF) of a degree $d$ is defined as the fraction of nodes that have
degree greater than the degree $d$. As in \cite{SigFal2003}, the
degree exponent, $\cal{D}$, is defined as the slope of CCDF$(d)$
against $d$ on a log-log  plot. An alternate definition of the degree
exponent is one based on the frequency $f(d)$ with which a node of
degree $d$ appears in the network. Either definition would serve the
purposes of this work but the degree exponent based on CCDF permits
statistically superior curve-fitting to determine the slope of the
log-log plot. 

\begin{figure}[!t]
\begin{center}
 \epsfig{file=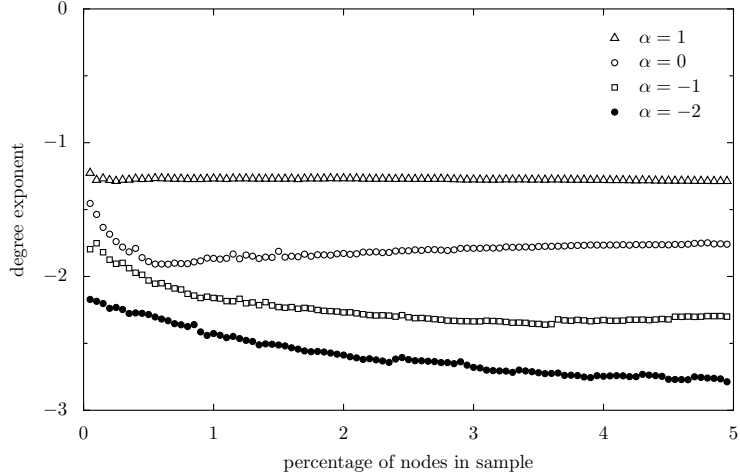,height=2.5in}
  \caption{The degree exponent of sample subgraphs extracted by the
    Biased Random Walk with Fly-Back (BRW-FB) for different values of
    $\alpha$. The plot shows that there is a linear relationship
    between $\alpha$ and the degree exponent to which the sample
    subgraph converges.} 
  \label{fig:BA2-1000000-linear-ccdfSlope}
\end{center}
\end{figure}

Given an estimate of the degree exponent of the full
graph $G$, our sampling algorithm is based on compensating for the
biases introduced by BFS-based methods such as Snowball and Forest
Fire. Our algorithm begins with a random node in the known part of the
graph $G$ and starts a biased random walk until it finds a new
unvisited node and then, flies back to the starting node to begin
another biased random walk. The bias in the random walk is
parametrized by $\alpha$ (we will shortly discuss how we determine
$\alpha$). Given $\alpha$, if the algorithm is at node $x$, then it
visits a neighbor $y$ of $x$ with probability $B(x,y)$ given by:
\begin{equation}
\label{bxy}
B(x,y) = \displaystyle{ \frac{[\mbox{degree}(y)]^\alpha}{\displaystyle \sum_{n \in \Gamma (x)}[\mbox{degree}(n)]^\alpha}}
\end{equation}
In other words, the walk proceeds to a neighbor $y$ with a probability
proportional to the degree of $y$ in $G$ raised to the power of
$\alpha$. Visited nodes are added to the sample and this continues
until the desired sample size is reached. The induced subgraph of
these nodes becomes the sample subgraph $S$.  Algorithm 1 presents the
pseudo-code of BRW-FB.

\begin{algorithm}
\caption{Biased Random Walk with Fly Back (BRW-FB)}
\textbf{Input:} $G$ (input graph), $h$ (desired sample size), $\alpha$ \\
\textbf{Output:} $S$ (sample graph)\\
\label{mine1}
\begin{algorithmic}
\STATE $N = \phi$ (set of nodes in sample graph)
\STATE Choose random node $p$ in $G$
\STATE Add $p$ to $N$
\WHILE{$|N| < h$}
\STATE foundANewNode $\leftarrow$ False
\STATE $x \leftarrow p$
\WHILE{not foundANewNode}
\STATE Choose a neighbor $y$ of $x$ with probability $B(x,y)$ (see
Equation (\ref{bxy}))
\IF{$y \in N$}
\STATE $x \leftarrow y$
\ELSE
\STATE foundANewNode $\leftarrow$ True
\STATE Add $y$ to $N$
\ENDIF
\ENDWHILE
\ENDWHILE
\STATE $S \leftarrow $ subgraph of $G$ induced by node set $N$
\RETURN{$S$}
\end{algorithmic}
\end{algorithm}

Figure \ref{fig:BA2-1000000-linear-ccdfSlope} presents an empirical demonstration of the
linear relationship between $\alpha$ and the degree exponent of the
sample generated by BRW-FB. In this figure, we use values of $\alpha$
in the range between $-2$ and $1$ based on our ongoing work on
crawling the network of YouTube users and the network of web
pages. While this relationship is different on different networks, the
linearity of it persists across all networks on which we have
attempted our algorithm. If we know the degree exponent yielded by two
chosen values of $\alpha$, the linear equations are readily solved to
generate an estimate of the relationship between $\alpha$ and the
degree exponent for the graph under consideration. Given this linear
relationship, it is possible to target a specific degree exponent in
the sample subgraph with the choice of an appropriate $\alpha$. 

We now present our {\em Tiny Sample Extractor}, which first executes the
BRW-FB with $\alpha = 0$ and $\alpha = 1$ to estimate the sensitivity of the
degree exponent to $\alpha$ and determine the underlying
relationship. Extrapolating based on this linear relationship, the
Tiny Sample Extractor computes the $\alpha$ corresponding to the
degree exponent estimated by the Metropolized Random Walk. Algorithm 2
presents the pseudo-code.

\begin{algorithm}
\caption{Tiny Sample Extractor}
\textbf{Input:} $G$, $h$ (desired sample size)\\
\textbf{Output:} $S$ (sample graph)\\
\label{mine2}
\begin{algorithmic}
\STATE $D \leftarrow $ MRW$( G, h )$
\STATE $S_0 \leftarrow $ BRW-FB$( G, h, 0 )$
\STATE $D_0 \leftarrow $ degree exponent of $S_0$
\STATE $S_1 \leftarrow $ BRW-FB$( G, h, -1 )$
\STATE $D_1 \leftarrow $ degree exponent of $S_1$
\STATE
\STATE $\alpha \leftarrow \displaystyle{-\left( \frac{D-D_0}{D_1-D_0}
  \right) }$
\STATE
\STATE $S \leftarrow $ BRW-FB$( G, h, \alpha )$
\RETURN{$S$}
\end{algorithmic}
\end{algorithm}

\section{Performance Analysis}
\label{sec:results}

We use the Barab\'{a}si-Albert scale-free network described in
\cite{BarAlb1999} for purposes of comparison. This being an extremely
well-behaved network, it illustrates more acutely the fact that the
existing sampling strategies do not approach properties of the actual
network even with one-fifth of the nodes in the sample. The specific
instance of the Barab\'{a}si-Albert network we choose is one that
begins with two unconnected nodes. When each new node is 
added to the network, two edges are created between it and two
pre-existing nodes. The probability with which an edge connects to an
existing node of degree $d$ is $d/\sum d_i$ where $d_i$ is the degree
of node $i$. The resulting network is a connected network.

\begin{figure*}[!t]
\begin{center}
    \subfigure[{Metropolized Random Walk}]{
       \label{fig:BA2-1000000-MRW3-20p-ccdfSlope}
       \epsfig{file=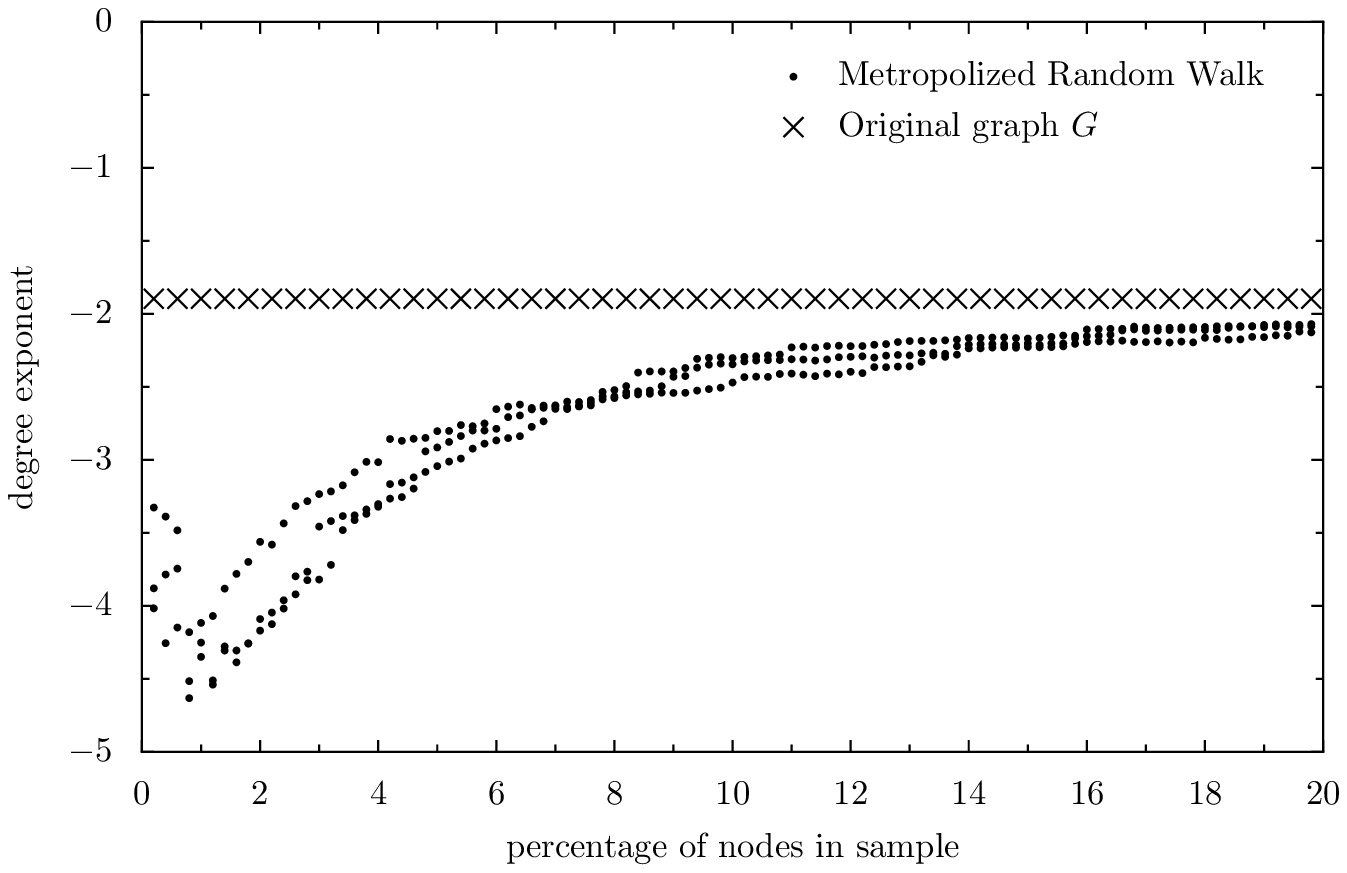,height=1.8in}
       }
     \qquad
    \subfigure[{Snowball sampling}]{
        \label{fig:BA2-1000000-SB100-20p-ccdfSlope}
        \epsfig{file=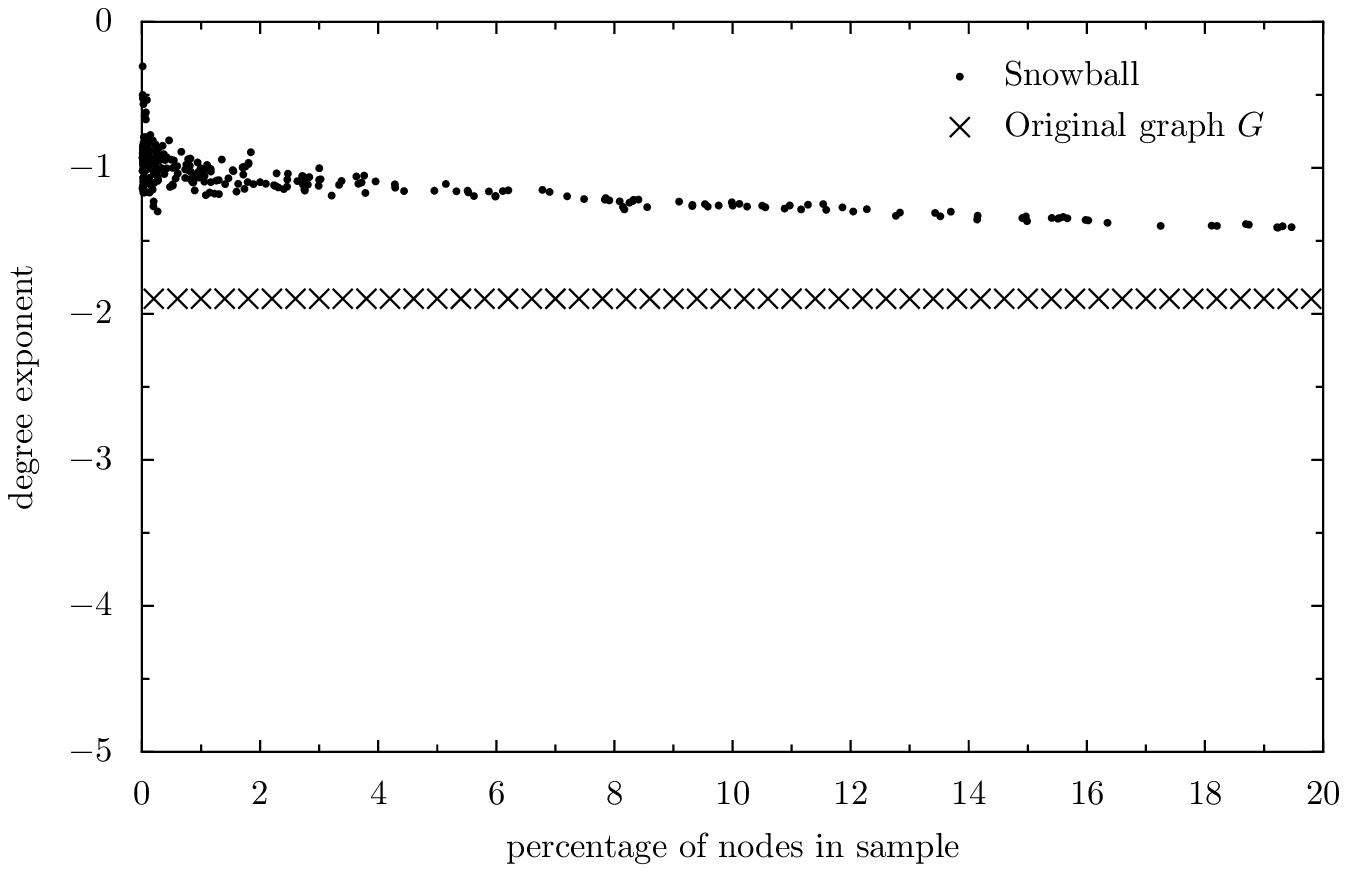,height=1.8in}
        }\\
    \subfigure[{Forest Fire}]{
        \label{fig:BA2-1000000-FF100-20p-ccdfSlope}
        \epsfig{file=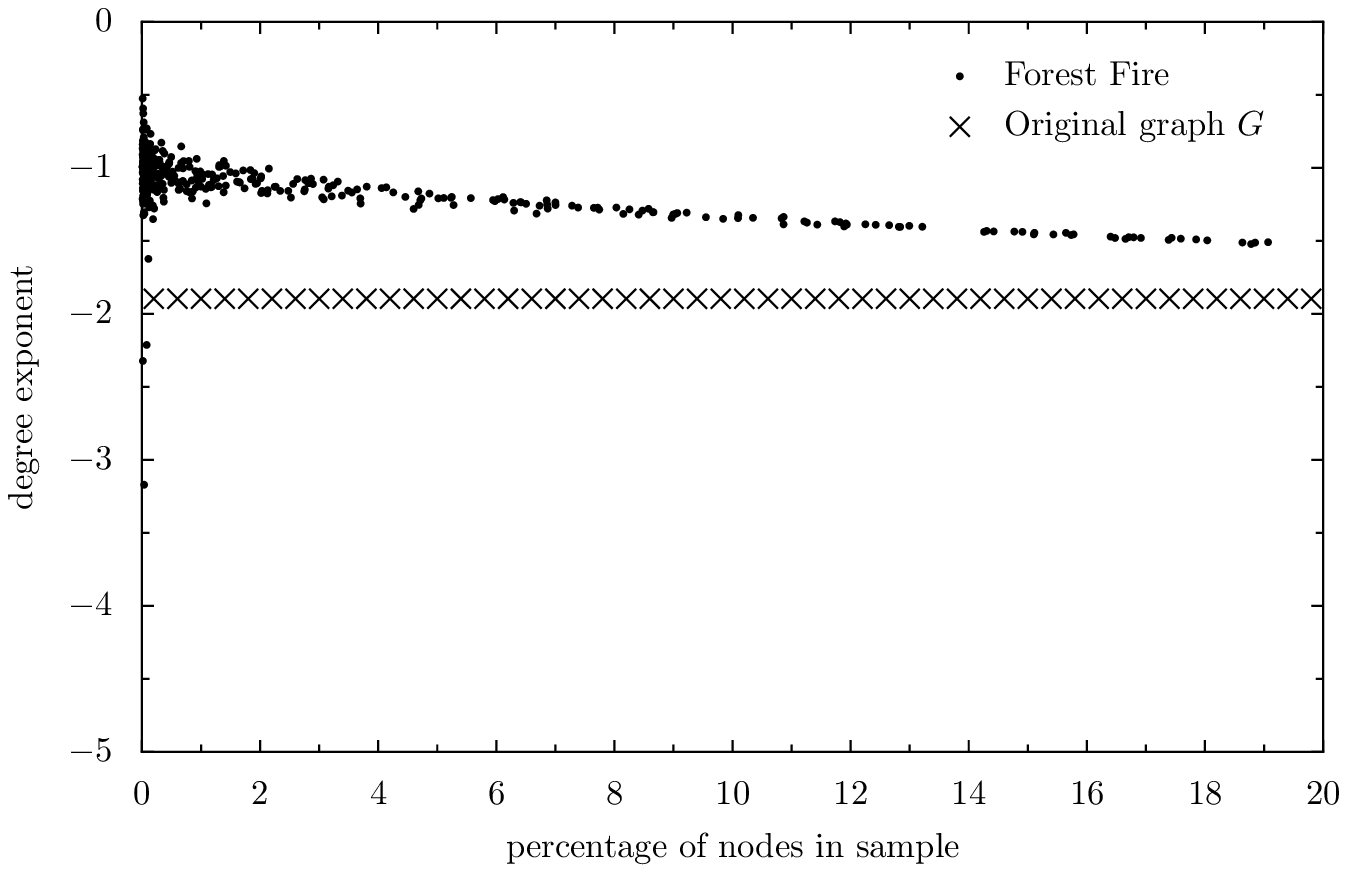,height=1.8in}
        }
     \hskip 0.3in
    \subfigure[{Tiny Sample Extractor}]{
        \label{fig:BA2-1000000-BRW3-20p-ccdfSlope}
        \epsfig{file=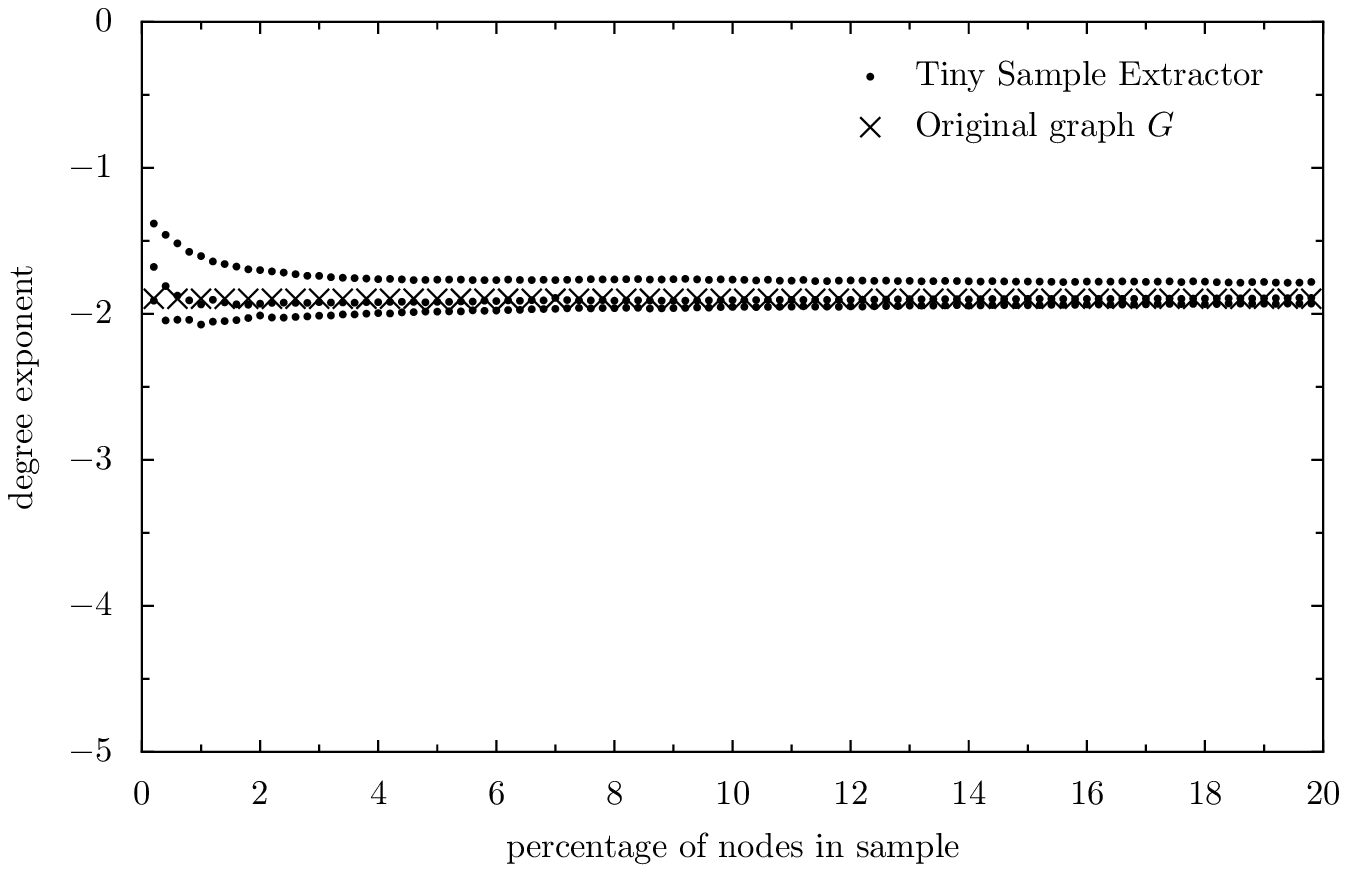,height=1.8in}
        }
    \caption{The degree exponent of sample subgraphs obtained using different sampling
      strategies. The original graph used is a million-node
      Barab\'{a}si-Albert scale-free graph with two edges added per new
      node. For the Metropolized Random Walk and the Tiny Sample
      Extractor, we extract three sample subgraphs. For Snowball
      sampling and Forest Fire, we extract hundred sample graphs. } 
\label{fig:ccdfSlopes}
\end{center}
\end{figure*}

The degree distributions of the samples extracted from this network by
four sampling strategies are shown in Figure \ref{fig:ccdfSlopes}. 
Figure \ref{fig:BA2-1000000-MRW3-20p-ccdfSlope} plots the degree
exponents of the subgraphs induced by the nodes visited by the
Metropolized Random Walk (MRW) as the set of nodes in the sample
grows. The figure plots three different samples, each starting from a
different random node. The goal of MRW is {\em node sampling} and not
{\em graph sampling}; therefore, it is not quite fair to the MRW
algorithm to plot the degree distribution in the induced
subgraph. However, we do so here largely to illustrate that node
sampling does not directly yield a representative subgraph.

Figure \ref{fig:BA2-1000000-SB100-20p-ccdfSlope} plots the degree
exponents achieved by Snowball sampling. The subgraph generated by
Snowball sampling grows very fast because of its BFS approach, and
therefore, very soon includes a large percentage of the
nodes. Step $i$ of the sampling strategy includes all nodes reachable
in $i$ hops or less from the starting node. Since only a few (2--4)
steps is needed to reach 20\% or more of the nodes in the network,
plotting degree exponents for only a small number of samples as they
grow does not fully illustrate the rate of convergence of the subgraph
to the degree distribution of the original graph. The figure,
therefore, plots degree exponents from one hundred samples as each sample
grows to 20\% of the network. As mentioned in Section
\ref{sec:related}, Snowball sampling over-samples high-degree nodes
and, as a result, the induced subgraph does not quite approach the degree
exponent of the original graph. In fact, even with 20\% of the nodes
in the sample, the degree exponent of the sample subgraph does not
converge to that of the original graph. Figure
\ref{fig:BA2-1000000-FF100-20p-ccdfSlope} similarly plots the degree
exponents corresponding to one hundred different subgraphs extracted by
the Forest Fire sampling strategy. As is readily observed, its
performance is very similar to Snowball sampling, though slightly
better.

Finally, Figure \ref{fig:BA2-1000000-BRW3-20p-ccdfSlope} plots the
degree exponents reached by the Tiny Sample Extractor. As in the case
of the Metropolized Random Walk, we plot the degree exponents for
three samples. The figure demonstrates that the Tiny Sample Extractor
converges to the degree distribution of the original graph
significantly faster than other sampling algorithms. 

We now focus on two additional properties of graphs: the average clustering
co-efficient and its assortativity \cite{ChaFal2006}.
Since the Forest Fire and Snowball sampling strategies are similar in
performance with the Forest Fire faring slightly better, for purposes
of clarity in the plots, we omit Snowball sampling in subsequent analysis. 

\begin{figure*}[!t]
\begin{center}
    \subfigure[{Assortativity}]{
       \label{fig:BA2-1000000-Assortativity}
       \epsfig{file=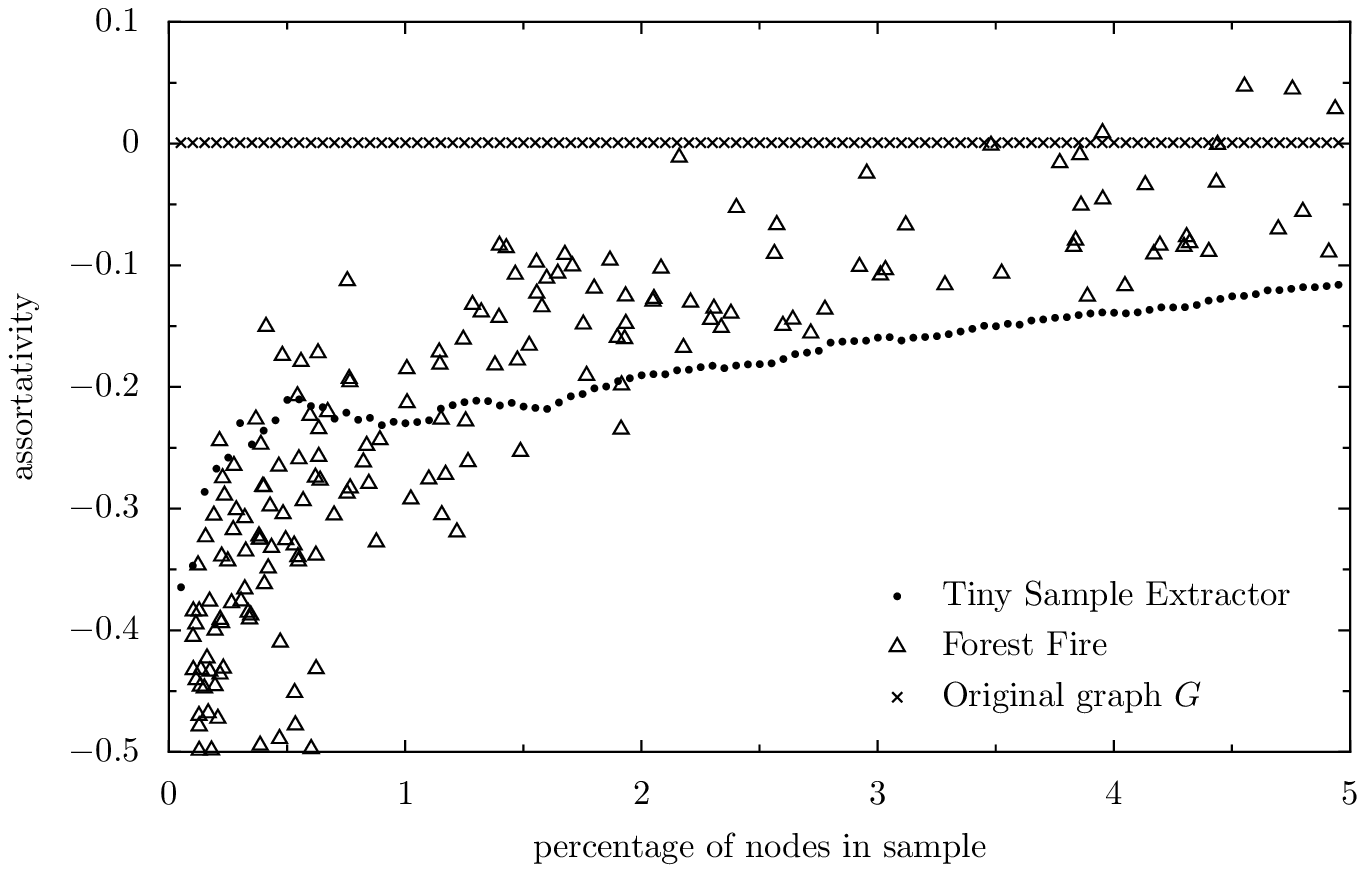,height=1.8in}
       }
     \hskip 0.3in
    \subfigure[{Average clustering co-efficient}]{
        \label{fig:BA2-1000000-ClusteringCoeff}
        \epsfig{file=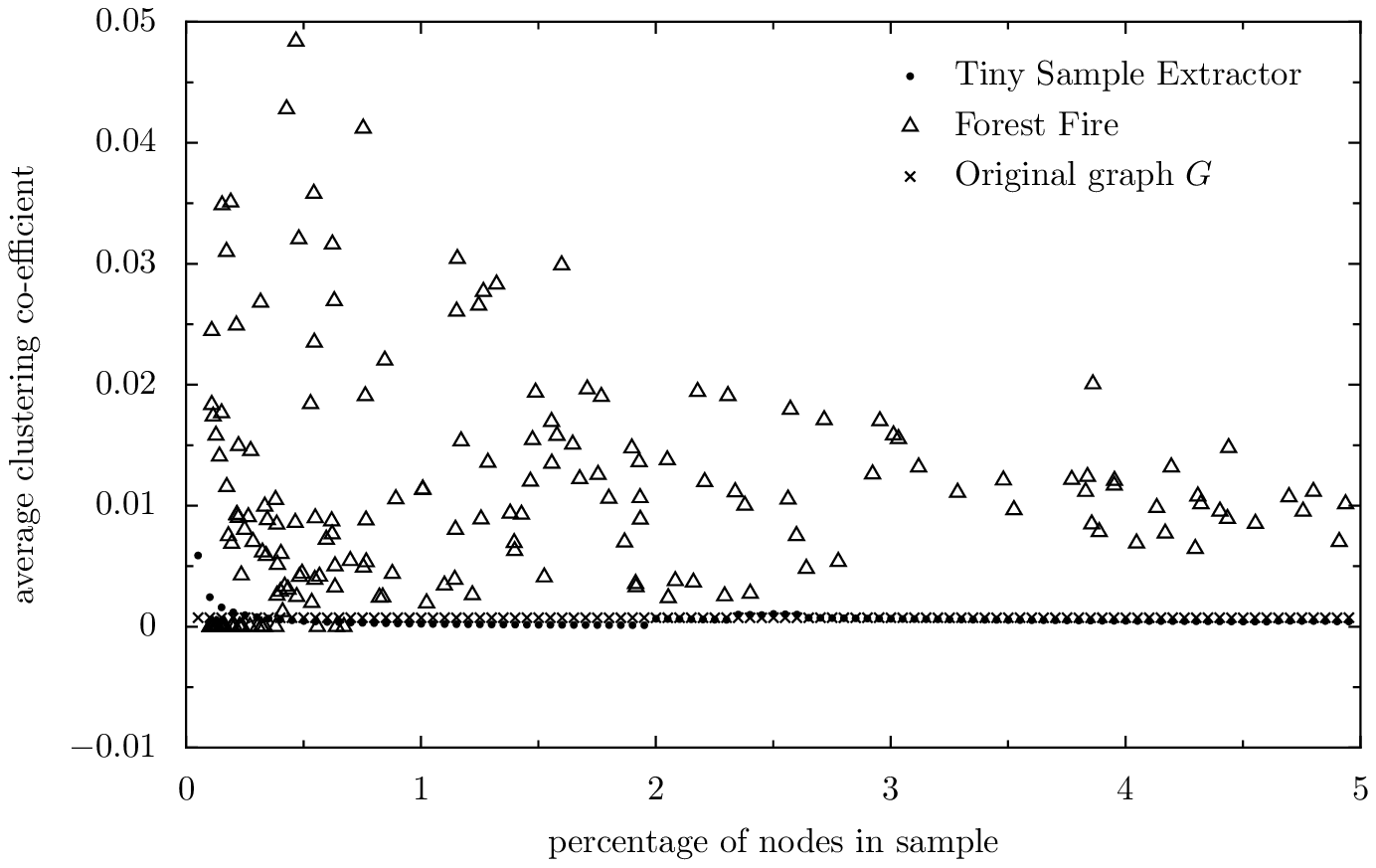,height=1.8in}
        }
   \caption{A comparison of assortativity and clustering properties of
     the sample subgraphs in relation to those of the original graph.}
\label{fig:BA-other}
\end{center}
\end{figure*}

Figure \ref{fig:BA2-1000000-Assortativity} plots the assortativity of the
sample graphs generated by Forest Fire and the Tiny Sample
Extractor. Assortativity measures the tendency of nodes to attach to
other nodes that are similar or different in any particular way. The
most commonly used definition of ``similarity'' used in studying
assortativity is one based on degrees. We define assortativity as:
\[
{\displaystyle \frac{ \langle d_id_j \rangle - \langle d_i \rangle
    \langle d_j \rangle }{\sqrt{ ( \langle d_i^2 \rangle - \langle d_i
      \rangle ^2
    ) ( \langle d_j^2 \rangle  - \langle d_j \rangle ^2 )} } }
\]
where $d_i$ and $d_j$ are degrees of nodes at either end of an edge
and the $\langle . \rangle$ notation represents an average over all edges in the
network. The performance of Forest Fire is not as well-behaved as
that of the Tiny Sample Extractor but slightly closer to the
assortativity of the original graph. 

Figure \ref{fig:BA2-1000000-ClusteringCoeff} plots the average clustering 
co-efficient of the sample graphs and the original graph. We define
the clustering co-efficient of a node $v$ as:
\[
\displaystyle{ \frac{2T(v)}{d_v(d_v - 1)} }
\]
where $T(v)$ is the number of triangles that exist through node
$v$. Note that the clustering co-efficient measures the fraction of
triangles that exist out of all potential triangles through a
node. The average clustering co-efficient is the average over all
nodes. We find that the Tiny Sample Extractor is highly accurate in
generating a sample that matches the average clustering co-efficient
of the original graph. In fact, as far as the clustering co-efficient
is concerned, the algorithm converges to that of the original graph
with as little as one percent of the nodes in the sample.

\section{Concluding Remarks}
\label{sec:conclusion}

In this report, we have presented a new crawling algorithm, called the
Tiny Sample Extractor, which extracts a small sample subgraph from a
large graph while retaining its essential properties, in particular
its degree exponent and the clustering co-efficient. A key feature of
our algorithm is that it achieves a convergence to these properties of
the original graph faster with a smaller sample than other
algorithms. This allows a crawler to take multiple snapshots of the
crawled network in order to study the temporal evolution of large
dynamic networks.

However, this work also illustrates that the problem of extracting a
subgraph that is representative of the full graph with respect to all
its properties remains unsolved. As shown in Section
\ref{sec:results}, the Tiny Sample Extractor does not do well with
respect to the assortativity of the graphs, even though it does better
than other algorithms on degree distribution and the clustering
co-efficient.

\end{document}